\documentclass[10pt,aps,pra,twocolumn]{revtex4-2}
\usepackage[dvipsnames]{xcolor}
\usepackage[colorlinks,allcolors=Blue,linktocpage]{hyperref}
\usepackage{amsmath,amsfonts,wasysym}
\usepackage{newtxtext,newtxmath,mathrsfs}
\usepackage{graphicx,pict2e,multirow}
\usepackage{CJK}
\makeatother
\begin{document}
\begin{CJK}{UTF8}{gbsn}
    \title{Quantum simulation of two-dimensional $\mathrm{U}(1)$ gauge theory in Rydberg and Rydberg-dressed atom arrays}
    
    \author{Zheng Zhou (周正)$^{2,3,4}$}
    \author{Zheng Yan (严正)$^{5,6}$}
    \author{Changle Liu (刘长乐)$^{1,7}$}
    \email{liuchangle89@gmail.com}
    \author{Yan Chen (陈焱)$^{2,8}$}
    \author{Xue-Feng Zhang (张学锋)$^{7,9}$}
    
    \affiliation{$^1$School of Physics and Mechatronic Engineering, Guizhou Minzu University, Guiyang 550025, China}
    \affiliation{$^2$Department of Physics and State Key Laboratory of Surface Physics, Fudan University, Shanghai 200438, China}
    \affiliation{$^3$Perimeter Institute for Theoretical Physics, Waterloo, Ontario N2L 2Y5, Canada}
    \affiliation{$^4$Department of Physics and Astronomy, University of Waterloo, Waterloo, Ontario, Canada N2L 3G1}
    \affiliation{$^5$Department of Physics, School of Science and Research Center for Industries of the Future, Westlake University, Hangzhou 310030, China}
    \affiliation{$^6$Institute of Natural Sciences, Westlake Institute for Advanced Study, Hangzhou 310024, China}
    \affiliation{$^7$Department of Physics and Chongqing Key Laboratory for Strongly Coupled Physics, Chongqing University, Chongqing 401331, China}
    \affiliation{$^8$Collaborative Innovation Center of Advanced Microstructures, Nanjing 210093, China}
    \affiliation{$^9$Center of Quantum Materials and Devices, Chongqing University, Chongqing 401331, China}
    \date{\today}
    
    \begin{abstract}
        Simulating $\mathrm{U}(1)$ quantum gauge theories with spatial dimension greater than one is of great physical significance. Here we propose a simple realization of $\mathrm{U}(1)$ gauge theory with Rydberg and Rydberg-dressed atom arrays. Within the experimentally accessible range, we find that the various aspects of the $\mathrm{U}(1)$ gauge theory can be well simulated, such as the emergence of topological sectors, incommensurability, and the Rokhsar-Kivelson point that hosts deconfined charge excitations and degenerate topological sectors. Our proposal is promising to implement experimentally and exhibits pronounced quantum dynamics. 
    \end{abstract}
    
    
    \maketitle
    \end{CJK}
    
    \emph{Introduction.}~---~Quantum simulations based on programmable Rydberg atom arrays have recently emerged as a powerful platform to explore exotic many-body physics. Rydberg simulators allow flexible design of lattice geometry~\cite{Rydberg_arrange1,Rydberg_arrange2}, and high tunability of interaction strength far beyond condensed matter experiments~\cite{Rydberg_array,Rydberg_review20}. A key element of quantum many-body simulations in Rydberg arrays is the Rydberg blockade mechanism, i.e. the van der Waals interactions strongly suppress the double occupancy of atomic Rydberg excitations within a certain blockade radius $R_c$~\cite{jaksch2000fast,lukin2001dipole,Rydberg_blockade}. In particular, the Rydberg blockade can lead to low-energy local constraints, which is an essential ingredient to realize lattice gauge theories~(LGT)~\cite{lgt_review,yan2022triangular}, a host of exotic phenomena such as quantum spin liquids~\cite{Balents,BalentsSavary,zhou2017quantum}, fractionalized excitations with Abelian and non-Abelian statistics~\cite{kitaev_anyon,Grover_nonabelian}. Very recently, significant progress has been made realizing two-dimensional $\mathbb Z_2$ LGT~\cite{Rydberg_Z2_1,Rydberg_Z2_2,Rydberg_Z2_3,Rydberg_Z2_4,semeghini2021probing}. In comparison, $\mathrm{U}(1)$ LGT involve more exotic phenomena like large number of topological sectors~\cite{kogut1979introduction,qdm_review}, incommensurability~\cite{rk_2dbp,ashvin1,zhou_incomm}, and deconfined quantum criticality~\cite{qdm_review}%,senthil2004deconfined,ashvin1,,wang2017deconfined}. 
    Moreover, it sheds light on simulating more complex gauge fields in high-energy physics \cite{QS_high,Rydberg_SU2}. Recently, the one-dimensional $\mathrm{U}(1)$ LGT has been successfully simulated in Rydberg platforms~\cite{Rydberg_1dU1_1,Rydberg_1dU1_2,Rydberg_1dU1_3,Rydberg_1dU1_4,Rydberg_1dU1_5}. Nevertheless, its implementation in higher dimensions involves complicated theoretical proposals~\cite{zoller,Rydberg_U1_1,Rydberg_U1_2,Rydberg_U1_3,Rydberg_U1_4}, and has not been achieved until recently. 
    
    One natural implementation of $\mathrm{U}(1)$ LGT is the quantum dimer model~(QDM) on the honeycomb lattice which can be mapped to the antiferromagnetic transverse field Ising model~(TFIM) on triangular lattice~\cite{qdm_review,LGT_QDM,frus_ising_2,qdm_honey,yan_mixed,ran2022fully,zhou_string,zhou_incomm,yan2021sweeping}. However, previous experiments simulating TFIM on Rydberg arrays~\cite{Rydberg_nature2} did not exhibit clear features of the $\mathrm{U}(1)$ gauge field. The physics of the `order-by-disorder'~(OBD)~\cite{OBD1,moessner_2000,moessner01,frus_ising_1,TFIM_DMRG} region with possible gauge structures are still in the mist of interplay between intertwined interactions.
    
    \begin{figure}[t]
        \centering
        \includegraphics[width=\linewidth]{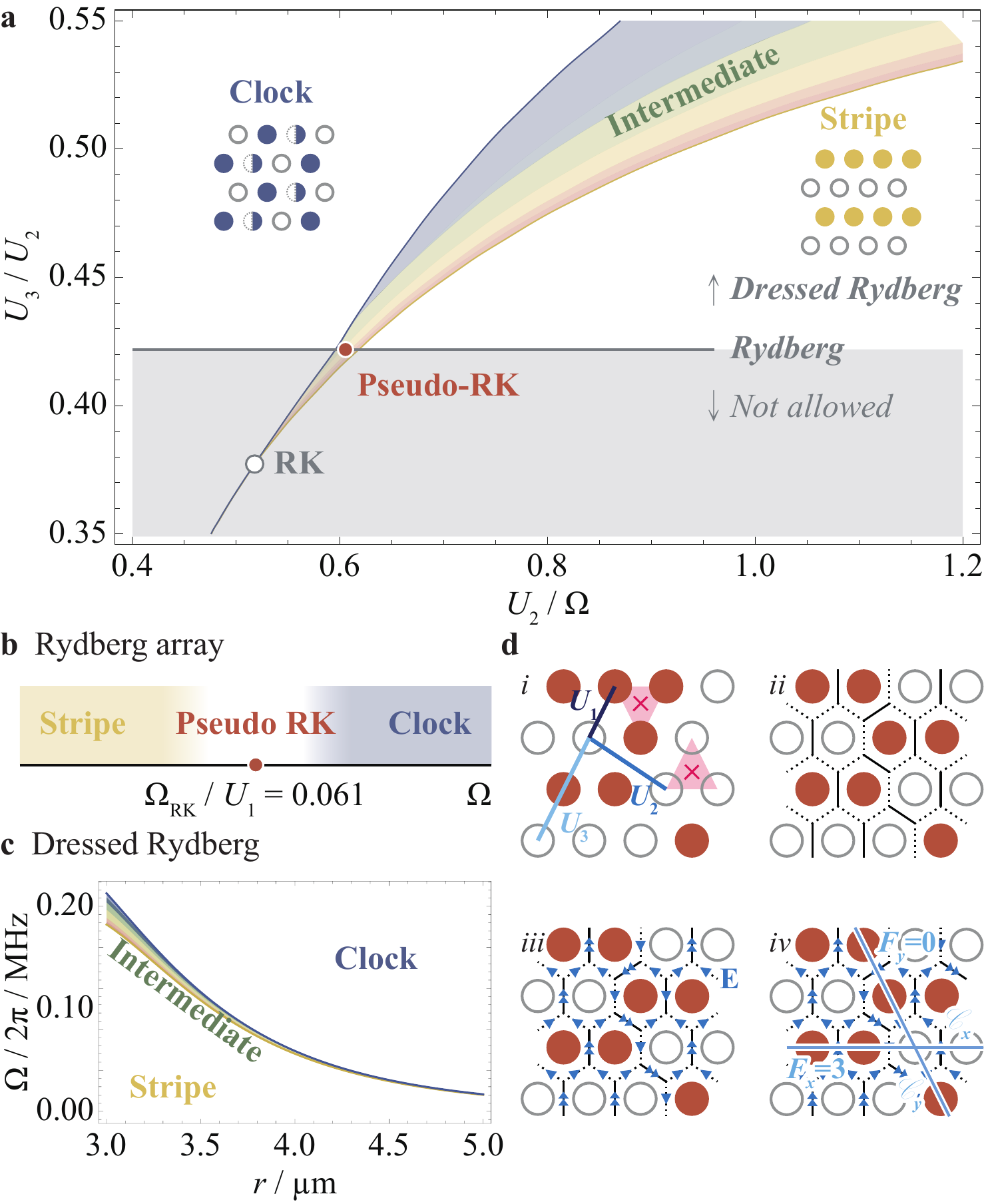}
        \caption{Phase diagram of the model at half filling. \textbf{a}, phase diagram on the $U_3/U_2$ plane; the dark gray line denotes the parameters accessed by the undressed Rydberg array and is further illustrated in \textbf{b}; the unshaded region denotes the accessible region for Rydberg-dressed arrays.
        The data are simulated with lattice length $L=24$ and $\Omega/U_1=0.2$. \textbf{c}, Phase diagram with Rydberg-dressed settings \cite{note_supp}; Here we take the experimental parameters of Ref.~\cite{Rydberg_nature2} with laser Rabi frequency $\Omega_L/2\pi=4.3\ \mathrm{MHz}$ and detuning $\delta_L/2\pi=1.3\ \textrm{MHz}$ and $c_6/2\pi=42729.4\ \mathrm{MHz\:\mu m^6}$. \textbf{d}, Illustration of the emergent $\mathrm{U}(1)$ gauge structure:
        \textit{i}, Interactions and the `triangle-rule' constraint. Constraint-violating triangles are marked shaded with crossings; 
        \textit{ii}, Mapping of atom configurations to fully-packed dimers and \textit{iii}, charge-free gauge field; and 
        \textit{iv}, Emergent topological sectors.
        }
        \label{fig_1}
    \end{figure}
    
    In this Letter, we propose a simple realization of $\mathrm{U}(1)$ LGT in Rydberg atom arrays with dressing technique \cite{Rydberg_dress,Rydberg_dress2,Rydberg_dress3,Rydberg_dress_Bloch}. We analyze the effective Hamiltonian with large-scale quantum Monte Carlo simulations. The softened repulsive interactions between Rydberg-dressed atoms provide various aspects of $\mathrm{U}(1)$ LGT, including emergent topological sectors, incommensurability, and the Rokhsar-Kivelson~(RK) point that hosts deconfined criticality. Our work demonstrates that the mystery of the OBD region can be uncovered by dressing Rydberg atoms, and many characteristic properties of the $\mathrm{U}(1)$ LGT could be experimentally observed.
    
    \emph{Model of Rydberg-dressed atoms.}~---~We consider that neutral atoms are arranged on a triangular lattice array \cite{Rydberg_nature2}. At low energies, each atom is regarded as a two-level system where $|\downarrow\rangle$ is the atomic ground state and $|\uparrow\rangle$ is the excited Rydberg state dressed with another hyperfine level~\cite{Rydberg_dress,Rydberg_dress_Bloch}. The system is described by the following Hamiltonian
    \begin{eqnarray}
    \nonumber
        H&=&\sum_i\left[-\frac{\Omega}{2}\left(|\uparrow\rangle\langle \downarrow|_i+|\downarrow\rangle\langle \uparrow|_i\right)-\delta|\uparrow\rangle\langle \uparrow|_i\right]\\ 
        &&+\sum_{ij}U_{ij}|\uparrow\rangle\langle \uparrow|_i\otimes|\uparrow\rangle\langle \uparrow|_j,
        \label{eq:jij_rydberg}
    \end{eqnarray}
    where $\Omega$ is the microwave Raman-Rabi frequency, $\delta$ is the detuning, and $U_{ij}$ is the repulsive interaction between Rydberg-dressed states. For undressed systems, the interactions are of the van der Waals type $U_{ij}=c_6/r_{ij}^{6}$. Upon dressing the interaction profile becomes softer, with slower decays at short distances but the long-distance behavior stays the same~\cite{note_supp,Rydberg_dress,Rydberg_dress3,Rydberg_dress_Bloch}. As will be discussed later, the $\mathrm{U}(1)$ gauge mapping requires the nearest-neighbor~(NN) coupling $U_1$ to be dominant, so we only focus on small $|\Omega|/U_1$ case. Meanwhile, the dressing technique renders the ratio between second- and third-NN interaction $U_2/U_3$ tunable, so that rich phenomena of the $\mathrm{U}(1)$ gauge field can emerge.
    
    Another requirement of the $\mathrm{U}(1)$ LGT is the half-filling of Rydberg-dressed states, which can be achieved by tuning the chemical potential $\delta$. At $\delta=\frac{1}{2}\sum_{j\neq i} U_{ij}$, the Hamiltonian exhibit an additional $\mathbb{Z}_2$-symmetry $\mathscr{P}$ that exchanges the Rydberg-dressed state and the ground state $\mathscr{P}|\uparrow\rangle_i =|\downarrow\rangle_i$, $\mathscr{P}|\downarrow\rangle_i =|\uparrow\rangle_i$, so that the half-filling condition is fulfilled.
    
    \emph{Phase diagram.}~---~We numerically investigate the model Eq.~\eqref{eq:jij_rydberg} with large-scale quantum Monte Carlo simulations~\cite{qmc_1,qmc_2,qmc_3,note_supp}. As the interaction is negligible at long distances, we truncate the interaction at the third-NN and study the phase diagram on the $U_2$-$U_3$ plane. The simulations are performed on $L\times L$ lattices with periodic boundary conditions at temperature $T=L^{-2}U_1$. 
    Most of our simulation takes $L=24,36$, and in the finite-size scaling we take $L$ from 12 to 36, as shown in Supplementary Materials (SM) \cite{note_supp}.
    
    We find a rich quantum phase diagram as shown in Fig.~\ref{fig_1}\textbf{a}. The clock and stripe orders are stabilized in the upper-left and lower-right regions in the phase diagram, respectively. The clock and stripe phases are separated by a first-order line at small $U_2$, and a fan-shaped intermediate region with larger $U_2$. The first-order line and the intermediate region terminate at a multicritical point. Through finite-size scaling, we determine its position in the thermodynamic limit $U_{2c}/\Omega = 0.547(5)$ and $U_{3c}/\Omega = 0.215(3)$.
    
    %Experimental proposal
    %\notexf{The quantum phase diagram can be presented with achievable specific experiment parameters. The triangular lattice Rydberg atom array is already realized with $^{87}$Rb atoms \cite{Rydberg_nature2}. Therefore, we can prepare the Rydberg dressed state $|\uparrow\rangle$ by coupling hyperfine state $|g\rangle=|5S_{1/2},F=2,m_F=2\rangle$ and Rydberg state $|r\rangle=|75S_{1/2},m_J=1/2\rangle$ via a two-photon process. The corresponding Rabi frequency $\Omega_L$ and detuning $\delta_L$ can be used to tune the soft-core potential between Rydberg-dressed atoms \cite{Rydberg_dress_Bloch}. Then, the ground state can choose another hyperfine state $|\downarrow\rangle=|5S_{1/2},F=1,m_F=1\rangle$. As shown in Fig.\ref{fig_1}\textbf{c}, all the main features are experimentally achievable. The experimental proposal in detail is discussed in SM \cite{note_supp}.}
    The essential features of the phase diagram are achievable with experimentally relevant parameters (Fig.~\ref{fig_1}\textbf{a}-\textbf{c}). Here we employ an $^{87}$Rb atom array that has already been realized in the experiment~\cite{Rydberg_nature2}, and the phase diagram is shown in Fig. 1\textbf{b}. 
    To access a more extended region of the phase diagram Fig. 1\textbf{a}, we perform Rydberg dressing, where the Rydberg-dressed state $|\uparrow\rangle$ is prepared by coupling hyperfine state $|g\rangle=|5S_{1/2},F=2,m_{F}=2\rangle$ and Rydberg state $|r\rangle=|53S_{1/2},m_{J}=1/2\rangle$ via a two-photon process~\cite{Evered2023}. The corresponding Rabi frequency $\Omega_{L}$ and detuning $\delta_{L}$ can be used to tune the soft-core potential between Rydberg-dressed atoms~\cite{Rydberg_dress_Bloch}. Meanwhile, one can choose another hyperfine level $|\downarrow\rangle=|5S_{1/2},F=1,m_{F}=1\rangle$ as the ground state. The experimental phase diagram is shown in Fig. 1\textbf{c} and more details are discussed in SM~\cite{note_supp}.
    
    \emph{Emergent $\mathrm{U}(1)$ gauge field and topological sectors}.~---~The structure of the phase diagram resembles that of the QDM on the honeycomb lattice, which is described by a $\mathrm{U}(1)$ LGT. To better understand the phase diagram, we start with the limit where $U_1$ is the dominant energy scale $U_1\gg |\Omega|, |U_2|, |U_3|$. At half Rydberg filling, the Hamiltonian can be recast in the following form
    \begin{equation}
    H_{U_1}=\frac{U_{1}}{2}\sum_{\triangle}\left(\sum_{i\in\triangle}|\uparrow\rangle_{i}\langle \uparrow|-\frac{3}{2}\right)^{2}+\mathrm{const}.
    \label{eq:V1}
    \end{equation}
    It is clear that the ground state must satisfy a local constraint that either one or two occupations of Rydberg-dressed states within each unit triangle are allowed, while zero or three occupations are blockaded (Fig.~\ref{fig_1}\textbf{d},\textit{i}). Such local constraint at half-filling gives rise to extensive ground-state degeneracy. This local constraint within each triangle serves as the key ingredient of the emergent $\mathrm{U}(1)$ gauge field: we define the emergent electric field on the (oriented) link of the dual honeycomb lattice. The value of the electric field (from A to B sublattice) is assigned 2 if the triangular lattice bond crossing this link is frustrated, and is assigned $-1$ otherwise. Here we define a bond to be `frustrated' if the two sites connected to this bond are occupied by the same states. Then this local constraint dictates that each unit triangle consists of one and only one frustrated bond, which translates to the charge-free condition of the $\mathrm{U}(1)$ gauge field at each dual vertex $\mathbf v$ (see Fig.~\ref{fig_1}\textbf{d},\textit{iii}):
    \begin{equation}
    Q_\mathbf{v}\equiv(\operatorname{div}\mathbf E)_\mathbf{v}=0.
    \end{equation}
    Note that this electric field configuration is equivalent to fully-packed dimers, by assigning each $\mathbf E=2$ link with one dimer and the $\mathbf E=-1$ link with no dimer~ (Fig.~\ref{fig_1}\textbf{d},\textit{ii}). We also note that as long as the local constraints are imposed by a sufficiently large $U_1$, its value no longer affects the physics of the system.
    \begin{figure}[t]
    \centering 
    \includegraphics[width=\linewidth]{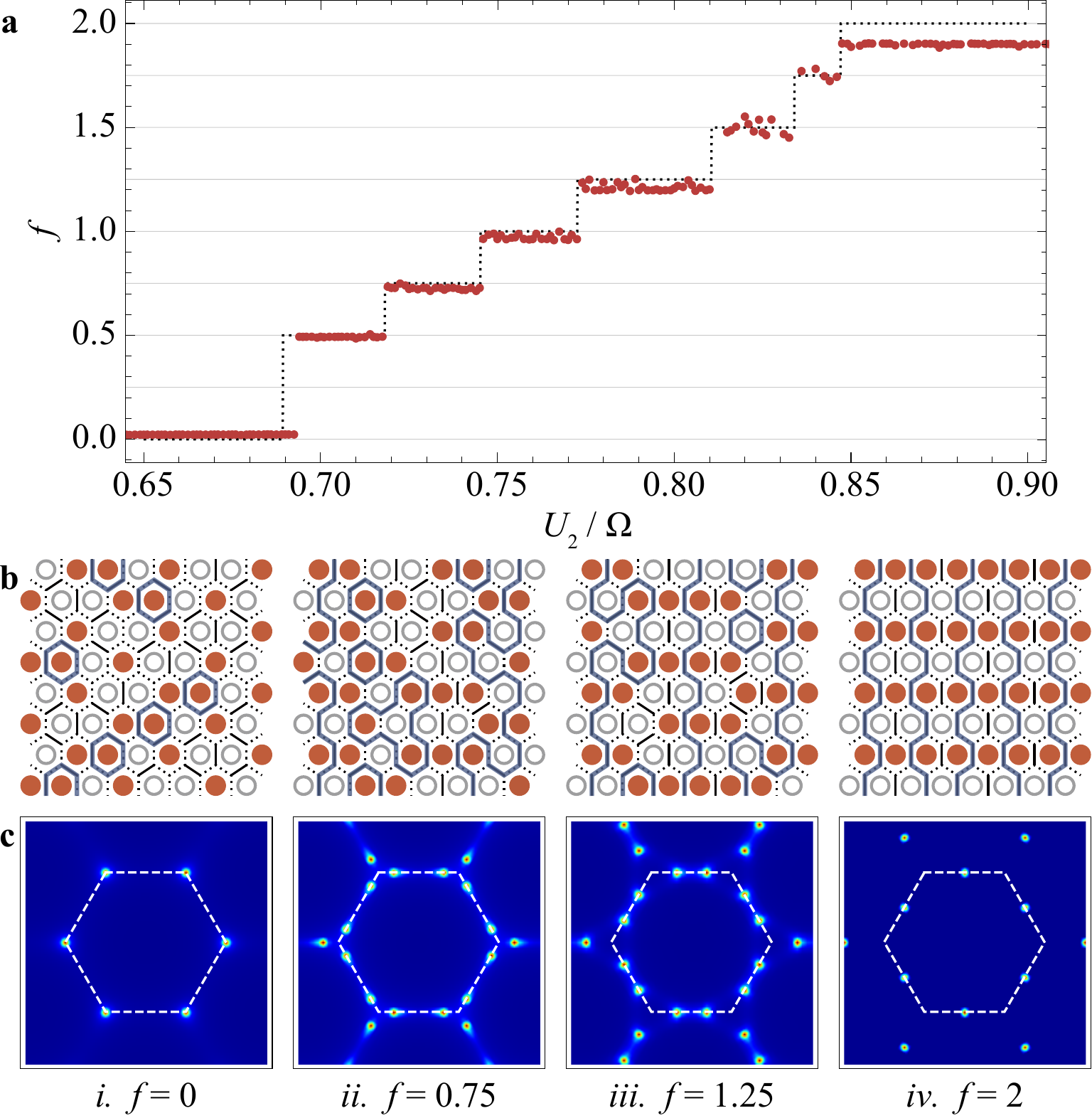} 
    \caption{Properties of each phase. \textbf{a}, Flux density as a function of $U_2/\Omega$. \textbf{b},\textbf{c}, Real space patterns (\textbf{b}) and Rydberg structure factors (\textbf{c}) at different topological sectors. The data are simulated along the $U_3/U_2=0.5$ line with $L=24$. }
    \label{fig_2} 
    \end{figure}
    
    A direct consequence of the $\mathrm{U}(1)$ gauge structure is the emergence of topological sectors, which can be defined as follows~\cite{qdm_honey}. For the two cuts $\mathscr C_x$ and $\mathscr C_y$ throughout the lattice (illustrated in Fig.~\ref{fig_1}\textbf{d},\textit{iv}), the corresponding electric fluxes that penetrate through these cuts are conserved quantities upon local fluctuations. Therefore, the low-energy Hilbert space can be divided into different topological sectors labeled by such fluxes $(F_x,F_y)$. For the ground states, one of the fluxes is zero. Without loss of generality, we set $F_y=0$ and use the flux density $f\equiv F_x/L_x$ to label topological sectors.
    
    The emergent $\mathrm{U}(1)$ gauge structure in the large $U_1$ limit is robust against local perturbations such as $U_2$, $U_3$, and $\Omega$. As illustrated in Fig.~\ref{fig_1}\textbf{a}, various density orders at different topological sectors are stabilized as the ground states in the phase diagram. These density wave orders can be also characterized by the ordering wave vectors in the Rydberg structure factor:
    \begin{equation}
    \mathscr S(\mathbf{Q})=\frac{1}{N}\sum_{ij} \left\langle \left(n_i-\frac{1}{2}\right)\left(n_j-\frac{1}{2}\right)\right\rangle e^{i\mathbf{Q}\cdot (\mathbf{r}_i-\mathbf{r}_j)},
    \end{equation}
    where $n_i=|\uparrow\rangle_i\langle \uparrow|$ is the occupancy of the Rydberg-dressed atom at site $i$. The flux densities $f$, real-space configurations, and the Rydberg structure factors of each density order are illustrated in Fig.~\ref{fig_2}.
    
\begin{figure}[t]
    \centering 
    \includegraphics[width=\linewidth]{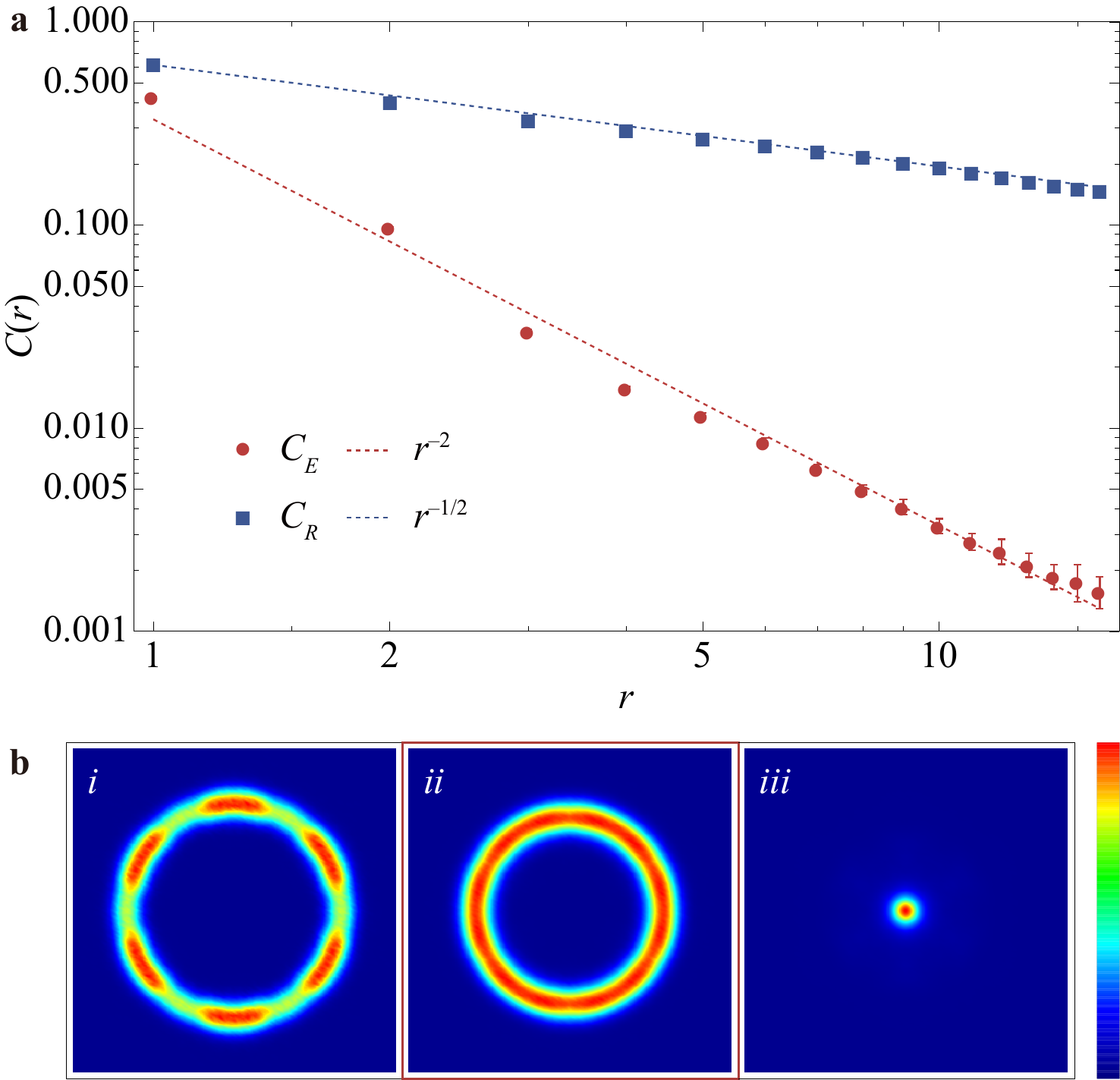} 
    \caption{\textbf{a}, Electric field and Rydberg density correlators at the multicritical point measured with $L=36$. Theoretical predicted behaviors are indicated by dashed lines. \textbf{b}, Order parameter histogram within the clock phase $U_2/\Omega=0.547,U_3/\Omega=0.1$ (\textit{i}), at the multicritical point $U_2/\Omega=0.547,U_3/\Omega=0.215$ (\textit{ii}) and in the stripe phase $U_2/\Omega=0.547,U_3/\Omega=0.4$ (\textit{iii}) measured with $L=24$.
    }
    \label{fig_3} 
    \end{figure}

    The three-sublattice `clock' state has the ordering wave vector at the K point and lives within the $f=0$ sector. The `clock' state is named after the six-fold clock anisotropy of the order parameter, as clearly illustrated in the histogram~( Fig.~\ref{fig_3}\textbf{b}\textit{i}). The clock state is stabilized via the quantum OBD mechanism~\cite{OBD1,moessner_2000,moessner01,frus_ising_1,TFIM_DMRG}. The two-sublattice `stripe' state is ordered at the M point with saturated flux density $f=2$. In between the `clock' and the `stripe' state lives a fan of intermediate phases with $0<f<2$. The ordering wave vector $\mathbf{Q}$ is typically incommensurate between the K and the M point, and is related with $f$ by $\mathbf{Q}=\left(\pm\frac{(2-f)\pi}{3},\pm\frac{2\pi}{\sqrt{3}}\right)$ and equivalent wave vectors connected by lattice symmetry.
    
    \emph{Multicritical RK point}.~---~By comparing the overall phase diagram with the generic QDM on the honeycomb lattice~\cite{ashvin1,qdm_honey}, we find that the most interesting aspect of our model is the correspondence between the multicritical point and the RK point of the QDM~\cite{rk_debut,qdm_review}. RK point is a quantum critical point separating clock and stripe states, and supports deconfined charge excitations right at the critical point~\cite{frus_ising_2,qdm_review}. It is described by $(2+2)$d quantum electrodynamics that supports quadratic gapless photons~\cite{note_supp,Henley,qdm_vbs}. On the honeycomb lattice, the RK point is multicritical with two relevant perturbations~\cite{rk_2dbp,ashvin1,qdm_honey}, hence can be reached by tuning two system parameters. Here we present concrete evidence that this multicritical point indeed corresponds to the RK point.
    
    A characteristic feature of the RK point is the exact ground state degeneracy among different topological sectors~\cite{Henley,qdm_review}. To verify, we measure the ground states with different sectors at the multicritical point with $L=24$, and find vanishing energy difference $\sim6\times10^{-4}L^{2}\Delta$. The tiny difference is accounted by the marginally and dangerously irrelevant terms~\cite{rk_2dbp,ashvin1}, which do not vanish at finite system size (details in SM ~\cite{note_supp}).
    
    Further evidence of RK point is provided by checking the order parameter correlations. We first measure the electric field correlator $C_{E}(\vec{\mathsf{R}}-\vec{\mathsf{R}}')=\langle\psi_E^{*}(\vec{\mathsf{R}})\psi_E^{\phantom{*} }(\vec{\mathsf{R}}')\rangle$, where 
    \begin{equation}
    \psi_E=E_{\mathbf{b}_1}+E_{\mathbf{b}_2}\mathrm{e}^{-\mathrm{i}2\pi/3}+E_{\mathbf{b}_3}\mathrm{e}^{\mathrm{i}2\pi/3}\label{VBS_OP}
    \end{equation}
    is the order parameter of the electric field. Here $E_{\mathbf{b}_i}$, $i=1,2,3$ represent the electric field along three directions. At RK point, such correlator is predicted to follow the long-distance behavior $C_{E}(\vec{\mathsf{R}})\sim|\vec{\mathsf{R}}|^{-2}$~\cite{note_supp,dimer_cor}, and is clearly observed in our simulations~(Fig.~\ref{fig_3}\textbf{a}, red line). We also measure the Rydberg density correlator $C_{R}(\vec{\mathsf{R}}-\vec{\mathsf{R}}')=\langle\psi_R^{*}(\vec{\mathsf{R}})\psi_R^{\phantom{*}}(\vec{\mathsf{R}}')\rangle$,
    where 
    
    \begin{equation}
    \psi_R=n_{A}+n_{B}\mathrm{e}^{\mathrm{i}2\pi/3}+n_{C}\mathrm{e}^{-\mathrm{i}2\pi/3}\label{eq:spin_OP}
    \end{equation}
    is the order parameter of the clock density-wave. The Rydberg correlator is predicted to satisfy $C_{R}(\vec{\mathsf{R}})\sim|\vec{\mathsf{R}}|^{-1/2}$~\cite{note_supp}, also consistent with our results~(Fig.~\ref{fig_3}\textbf{a}, blue line).
    
    A hallmark of deconfined $\mathrm{U}(1)$ gauge field is the presence of emergent continuous symmetries, which can be examined in histograms of the order parameters. We measure the histograms of the clock order parameter $\psi_R$ around the multicritical point, see Fig.~\ref{fig_3}\textbf{b}. In the clock phase, $\psi_R$ is pinned to six distinct values in the histogram, indicating relevance of instantons~\cite{Polyakov,Henley,rk_2dbp}~(Fig.~\ref{fig_3}\textbf{b}\textit{i}). For the stripe phase the histogram shrinks into a central peak~(Fig.~\ref{fig_3}\textbf{b}\textit{iii}) indicating absence of three-sublattice ordering. At the multicritical point, we find that the $\mathrm{U}(1)$ symmetry emerges as a symmetric ring in the histogram~(Fig.~\ref{fig_3}\textbf{b}\textit{ii}). The right presence of the emergent $\mathrm{U}(1)$ symmetry at the multicritical point is consistent with the irrelevance of instantons and deconfinement of charges~\cite{senthil2004deconfined,ashvin1,qdm_review}. Considering that the local density profile of Rydberg-dressed states can be directly detected in experiments~\cite{Rydberg_review20}, all the observables discussed above can be extracted.
    
\begin{figure}[b]
        \centering
        \includegraphics[width=\linewidth]{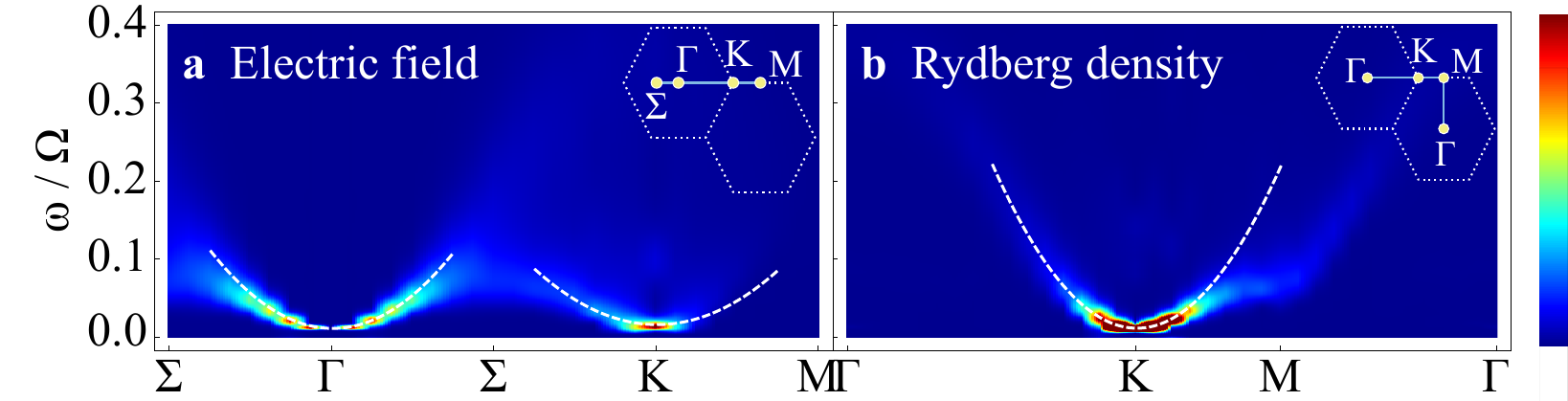}
        \caption{\textbf{a}, Electric field and \textbf{b}, Rydberg density excitation spectra at the multicritical point. The data are simulated at the $f=0$ topological sector with system size $L=24$. 
        }
        \label{fig_4}
    \end{figure}

    \emph{Excitations at the multicritical point.}---Here we turn to the dynamical properties at the multicritical point. We measure the dynamical electric field and Rydberg density correlators: 
    \begin{equation}
    \begin{aligned}G_{E}(\mathbf{q},\tau) & =\frac{1}{L^{2}}\sum_{\mathbf{R,R'}}e^{i(\mathbf{R}-\mathbf{R'})\cdot\mathbf{q}}\langle E_{\mathbf{R}}^y(\tau)E_{\mathbf{R'}}^y(0)\rangle,\\
    G_{R}(\mathbf{q},\tau) & =\frac{1}{L^{2}}\sum_{ij}e^{i(\mathbf{r}_{i}-\mathbf{r}_{j})\cdot\mathbf{q}}\langle n_{i}(\tau)n_{j}(0)\rangle.
    \end{aligned}
    \end{equation}
    Here $E_{\mathbf{R}}^y$ represents the electric field originating from site $\mathbf{R}$ along the $y$-direction. The spectra are obtained from the imaginary time correlations via stochastic analytical continuation~\cite{sac_1,sac_2,sac_3,note_supp}. 
    As widely discussed in the QDM, at RK point two gapless quadratic excitations present in the electric field spectra~\cite{rk_spec,rk_excitation}: the `resonon' around $\Gamma$ point that corresponds to the gauge photon, and the `pi0n' around $\mathrm{K}$ point due to the proximity to the three-sublattice ordered phase. These two quadratic excitations are indeed observed in our numerical spectra~(Fig.~\ref{fig_4}\textbf{a}). Further, the distinct curvatures of these two excitations~\cite{note_supp} exclude band folding and indicate their distinct physical nature.
    
    For the Rydberg density correlators, we observe only one branch of quadratic excitation stemming from the $\mathrm{K}$ point~(Fig.~\ref{fig_4}\textbf{b}), with curvature different  from the previous two. This excitation can be regarded as fractionalized `pi0n' that transforms non-trivially under the symmetry $\mathscr{P}$.
    Hence we dub this new excitation `pi0n{*}'. 
    
    \emph{Advantages of Rydberg-dressed platform.}~---~
    Our proposal has several advantages compared with previous ones realizing two-dimensional LGT. In our proposal, only one species of atom is required and the arrangement is the simple triangular lattice. Owing to the rapid development of Rydberg atom experiments, the Rydberg-dressed atom array on a triangular lattice is highly achievable in the near future. Moreover, our proposal overcomes an important issue on the lack of quantum dynamics in previous proposals. In LGT the quantum dynamics are contributed by the ring-exchange processes, which usually arise as high order perturbative terms hence strongly suppressed at low energies~\cite{Hermele,Savary,zoller,semeghini2021probing,Rydberg_Z2_1,Rydberg_Z2_2,Rydberg_Z2_3,Rydberg_Z2_4,Rydberg_U1_1,Rydberg_U1_2,Rydberg_U1_3,Rydberg_U1_4}. The suppressed quantum dynamics make the system glassy hence hard to be simulated experimentally~\cite{yan_glass}. In contrast, in our scheme, the gauge dynamics is directly proportional to the first order of the Rabi-frequency $\Omega$, which can be significant and highly tunable. We believe that our system provides an ideal platform for exploring exotic quantum aspects of LGT. 
    
    \emph{Discussions.}~---~Simulating LGT in more than one dimension is of great interest and physical significance. Our proposal based on Rydberg-dressed atom arrays provides a simple and direct realization of the two-dimensional $\mathrm{U}(1)$ LGT. It also allows further investigation of broad aspects of $\mathrm{U}(1)$ LGT. Moreover, it provides an ideal platform to investigate topological excitations such as `spinons' and `topological strings'~\cite{wan_string,zhou_string,zhou_incomm}. With rapid developments of the Rydberg platforms in recent years, various unconventional measurements such as snapshots and non-local correlation probes now become possible.
    
    The most interesting region in the phase diagram locates at the boundary between the clock and stripe states, which turns out to be most accessible in Rydberg-based experiments. For undressed Rydberg platforms, the experimentally accessible region is marked as the black thick line in Fig.~\ref{fig_1}\textbf{a}, which cuts through the intermediate region with negligible thickness. Such small thickness indicates that the relevant cubic anisotropy term is sufficiently small~\cite{ashvin1} so that the intermediate region lies sufficiently close to the celebrated RK point. As an evidence, the energy difference among topological sectors is $\sim 1.1\times10^{-3}\Omega L^2$, still much less than the typical energy scale of this system. Hence we claim that the RK point can be simulated directly with conventional Rydberg experiments. With Rydberg-dressed atoms~\cite{Rydberg_dress,note_supp} one can further access an extended region of the phase diagram (marked unshaded in Fig.~\ref{fig_1}\textbf{a}) that includes the intermediate phases. This region also hosts interesting physics, such as incommensurability, Kasteleyn transition~\cite{Kasteleyn_1,Kasteleyn_2,Kasteleyn_3}, and even the `Cantor deconfinement' of gauge field~\cite{rk_2dbp,ashvin1}. We expect that the interesting physics of this region will be unveiled with Rydberg-dressed atom arrays in future experiments.
    
    %\begin{acknowledgments}
        \textit{Acknowledgments.~---~}
        We wish to thank Rong Yu, Hai-Zhou Lu, Yuan Wan, Yin-Chen He, Yue Yu, Yiming Wang, Jiucai Wang and Chun-Jiong Huang for their valuable discussions. X.-F. Zhang thank Heng Shen for his practical suggestions on the feasibility of the experiment proposals. C.L. thanks Rong Yu for hospitality inviting him to visit Renmin University of China where part of the work is done. This work is supported by the National Key Research and Development Program of China (Grant No. 2022YFA1404204, 2022YFA1403700), the National Natural Science Foundation of China (Grants Nos.~12274086, 11534001 and 11925402). X.-F. Z. acknowledges funding from the National Science Foundation of China under Grants  No. 12274046, No. 11874094, No. 12147102 and No.12347101, Chongqing Natural Science Foundation under Grants No. CSTB2022NSCQ-JQX0018, Fundamental Research Funds for the Central Universities Grant No. 2021CDJZYJH-003, and Xiaomi Foundation / Xiaomi Young Talents Program. Z.Y. thanks the supports of the start-up funding of Westlake University. Z.Z. acknowledges support from the Natural Sciences and Engineering Research Council of Canada (NSERC) through Discovery Grants. Research at Perimeter Institute is supported in part by the Government of Canada through the Department of Innovation, Science and Industry Canada and by the Province of Ontario through the Ministry of Colleges and Universities.
    %\end{acknowledgments}
    
\let\oldaddcontentsline\addcontentsline% Store \addcontentsline
\renewcommand{\addcontentsline}[3]{}% Make \addcontentsline a no-op
%\bibliography{ref}

\let\addcontentsline\oldaddcontentsline% Restore \addcontentsline

\clearpage    
\onecolumngrid
\pagestyle{plain}
\fontsize{10pt}{15pt}\selectfont

\setcounter{page}{1}
\setcounter{equation}{0}
\setcounter{figure}{0}
\renewcommand{\thefigure}{S\arabic{figure}}
\renewcommand{\thetable}{S\Roman{table}}
\renewcommand{\thepage}{S\arabic{page}}
\renewcommand{\theequation}{S\arabic{equation}}
\renewcommand{\thesection}{\arabic{section}}

\begin{center}
    \noindent\large{\textbf{Supplemental Materials -- Quantum simulation of two-dimensional U(1) gauge theory in \\
    Rydberg and Rydberg-dressed atom arrays}}

    %\vspace{\baselineskip}
    %\noindent\large{\textbf{Supplemental Materials}}
\end{center}

\tableofcontents

\section{Experimental proposals}
In this section, we first derive the interaction profile between Rydberg-dressed atoms. Then we discuss some other experimental platforms that could be used in our work. At last, we discuss some potential challenges in experiments and how they could be overcome.

\begin{figure}[t!]
\centering \includegraphics[width=1\linewidth]{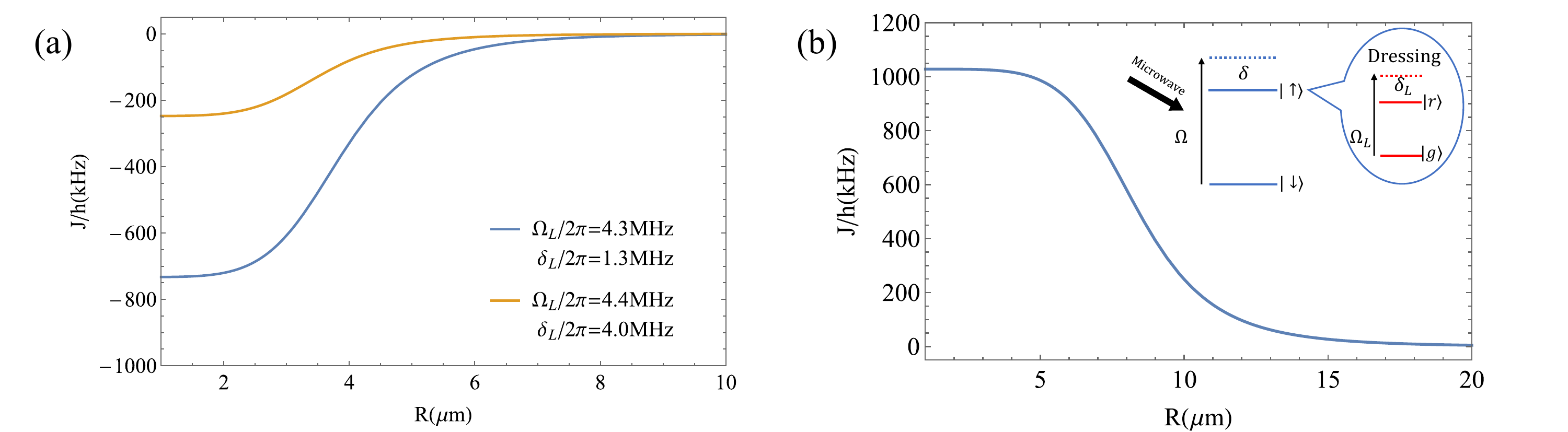} \caption{The soft-core EDDI potential $J(R)$ calculated by diagonalization of Eq.(\ref{hd}) for (a) $^{133}$Cs with experimental parameters same as Ref. \cite{Rydberg_dress} and (b) $^{87}$Rb with experimental parameters in the main text. Inset of (b): the transition processes related to the preparation of the Rydberg-dressed qubit. }
\label{fig:dress0} 
\end{figure}

A Rydberg-dressed state is made of coherent superposition between the Rydberg state $|r\rangle$ and the atomic ground state $|g\rangle$~\cite{Rydberg_dress3}. To derive the effective interaction between Rydberg-dressed states, we consider a two-site Rabi model:
%The preparation of the Rydberg-dressed atoms makes use of two states~\cite{Rydberg_dress3}, and the Hamiltonian is two-site Rabi model written as 
\begin{equation}
    H_{dd}=\sum_{i=1}^2 \left[\frac{\Omega_L}{2}\left(|g\rangle_i \langle r|+|r\rangle_i \langle g|\right) +\delta_L |r\rangle_i \langle r| \right] +V|r\rangle_1 \langle r| \otimes |r\rangle_2 \langle r|,
\end{equation}
where $\Omega_L$ is the Rabi frequency, $\delta_L$ is the detuning and $U=c_6/(r_1-r_2)^6$ is the electric dipole-dipole interaction (EDDI). 
%$\tau_i^\alpha$ ($\alpha=x,z$, $i=1,2$) are Pauli matrices acting on the basis of $|g\rangle$ and $|r\rangle$ at site $i$. 
Then, the matrix form of the Hamiltonian under the two-site basis $|gg\rangle$, $|gr\rangle$, $|rg\rangle$, and $|rr\rangle$ is 
$$H_{dd}=\left(
\begin{array}{cccc}
 0 & \frac{\Omega_L }{2} & \frac{\Omega_L }{2} & 0 \\
 \frac{\Omega_L }{2} & \delta_L  & 0 & \frac{\Omega_L }{2} \\
 \frac{\Omega_L }{2} & 0 & \delta_L  & \frac{\Omega_L }{2} \\
 0 & \frac{\Omega_L }{2} & \frac{\Omega_L }{2} & 2 \delta_L +U \\
\end{array}
\right).$$
The effective interaction between Rydberg-dressed atoms can be extracted from the energy levels of the two-site Rabi model.
After transforming to the new basis  $|gg\rangle$, ($|gr\rangle$+ $|rg\rangle$)/$\sqrt{2}$, ($|gr\rangle$- $|rg\rangle$)/$\sqrt{2}$, and $|rr\rangle$, we obtain
$$H'_{dd}=\left(
\begin{array}{cccc}
 0 & \frac{\Omega_L }{\sqrt{2}} & 0 & 0 \\
 \frac{\Omega_L }{\sqrt{2}} & \delta_L  & 0 & \frac{\Omega_L }{\sqrt{2}} \\
 0 & 0 & \delta_L  & 0 \\
 0 & \frac{\Omega_L }{\sqrt{2}} & 0 & 2 \delta_L +U \\
\end{array}
\right).$$
We can clearly find the singlet state ($|gr\rangle$- $|rg\rangle$)/$\sqrt{2}$ is decoupled with other states, so the Hamiltonian is changed into 
\begin{equation}
    H_{d}=\left(
\begin{array}{ccc}
 0 & \frac{\Omega_L }{\sqrt{2}} & 0 \\
 \frac{\Omega_L }{\sqrt{2}} & \delta_L  & \frac{\Omega_L }{\sqrt{2}} \\
 0 & \frac{\Omega_L }{\sqrt{2}}  & 2 \delta_L +U \\
\end{array}
\right).\label{hd}
\end{equation}
The ground state of $H_d$ is the Rydberg-dressed state, and its eigenvalue is the ground state energy or light shift~\cite{Rydberg_dress}. After subtracting the light shift of a lone atom $2\Delta E_{LS}^{(1)}=\delta_L-\sqrt{\delta_L^2+\Omega_L^2}$, the soft-core EDDI potential $J(R)$ can be obtained. 
%After extracting the coefficient of EDDI from Ref.~\cite{Rydberg_dress} $c_6/2\pi=18110\,\mu\textrm{m}^6\cdot$MHz, we can reproduce the Fig.~2(b) of Ref.~\cite{Rydberg_dress} and show it in Fig.~\ref{fig:dress0}(a). 
For the experimental proposal in the Ref.~\cite{Rydberg_dress}, we can reproduce the Fig.~2(b) of Ref.~\cite{Rydberg_dress} as shown in Fig.~\ref{fig:dress0}(a). 

From the above derivation, the reason that the interaction between Rydberg-dressed atoms is softened at small distances (see Fig.~\ref{fig:dress0}) becomes clear: with small inter-atomic distances where the EDDI $V$ dominates, the $|rr\rangle$ component that contributes to the Rydberg-dressed interaction is significantly suppressed at low energies. Therefore, we observe the saturation behavior of Rydberg-dressed interactions at small distances, instead of the diverging behavior of the bare EDDI $\sim~R^{-6}$. 

The Rydberg atom array with triangular lattice geometry has been experimentally realized \cite{Rydberg_nature2}, so our experimental proposal is based on Ref.~\cite{Rydberg_nature2} that also makes use of ultra-cold $^{87}$Rb atoms. 
%The ground state is $|g\rangle=|5S_{1/2},F=2,m_F=2\rangle$ and the Rydberg state is $|r\rangle=|75S_{1/2},m_J=1/2\rangle$. 
The ground state $|g\rangle=|5S_{1/2},F=2,m_F=2\rangle$ is the same as that in Ref.~\cite{Rydberg_nature2}. However, we choose a different Rydberg state $|r\rangle=|53S_{1/2},m_J=1/2\rangle$ in order to increase the light-matter interactions~\cite{Evered2023}. 
Owing to the $n^{11}$ dependence of the EDDI potential coefficient $c_6$, the corresponding $c_6$ for the $n=53$ Rydberg states can be inferred from Ref.~\cite{Rydberg_nature2} to be $c_6/2\pi\approx 42729.4\,\mu\textrm{m}^6\cdot$MHz. These two states can be coupled via a two-photon process with a 420-nm and 1013-nm laser, respectively~\cite{Evered2023}. 
Then, the effective Rabi frequency $\Omega_L$ and the detuning $\Delta_L$ can be tuned by adjusting the laser power and frequency, respectively. In this Letter, we set $\Omega_L/2\pi=5\ \mathrm{MHz}$ and $\delta_L/2\pi=1\ \textrm{MHz}$ which are experimental easily achievable \cite{Rydberg_dress_Bloch}, and the corresponding soft-core potential is shown in Fig.~\ref{fig:dress0}(b). After that, we take another hyperfine state $|\downarrow\rangle=|5S_{1/2},F=1,m_F=1\rangle$ and couple it to the dressed state $|\uparrow\rangle$ via a microwave \cite{Rydberg_dress_Bloch} as shown in inset of Fig.~\ref{fig:dress0} (b). The corresponding Raman-Rabi frequency $\Omega$ and the detuning $\delta$ can be tuned by changing the power and frequency of the microwave, respectively. The Raman-Rabi frequency $\Omega$ should be small compared with $U_1$ so that the strong coupling parameter regime associated with the emergent U(1) gauge structure can be reached. Meanwhile, it can not be too small, otherwise, the measurement or preparation of the state may require so long time that the system decoheres. As demonstrated in Ref.~\cite{Rydberg_dress_Bloch}, the microwave Rabi frequency $\Omega/2\pi$ can be set smaller as 12.5kHz which is definitely small enough for observing the exotic phenomena induced by the emergent U(1) lattice gauge field. 

Furthermore, the platform is not limited to the conventional $^{87}$Rb atoms in the Rydberg array. The development of distance-selective interaction may make the system more tunable~\cite{Rydberg_dress_bloch2}. Meanwhile, considering the preparation of the $^{133}$Cs Rydberg-dressed atoms can be implemented by only a single-photon transition~\cite{Rydberg_dress}, which can reduce the complexity of the corresponding experiments.

At last, we want to emphasize the temperature in the Rydberg array is not low in comparison with the optical lattice. Meanwhile, during the preparation of the Rydberg-dressed atom, the Rydberg-dressed states in different sites are not uniform due to the inhomogeneity of the experimental platform. However, these drawbacks may be overcome by employing the atom array-optical lattice hybrid system~\cite{op-array}.

%As shown in  of  (I. Bloch  NP 12 1095-1099 (2016)), the Rabi coupling . It is easy to increase $\Omega$. However, when it is too small, the time of measurement will be larger than the time of decoherence. %The dressed state $|\uparrow\rangle$ can be prepared by coupling Rydberg state $|r\rangle$ and the ground state $|g\rangle$, and the soft-core potential can be modified by changing both $\Omega_L$ and $\Delta_L$. %At last, we consider the experimental parameters mentioned before. The dressed state is prepared by taking one hyperfine state of $^{87}$Rb atom $|g\rangle=|5S_{1/2}, F=2,m_F=2\rangle$, and the Rydberg state is $|r\rangle=|75S_{1/2},m_J=1/2\rangle$.  As described in Fig.1 of the main text, the laser Rabi frequency and detuning are set to be $\Omega_L/2\pi=5\ \mathrm{MHz}$ and $\delta_L/2\pi=1\ \textrm{MHz}$, After 

\section{Height field representation}

In this section, we give an introduction to the height field representation
of the two-dimensional U(1) gauge theories in the context of bipartite dimer model. The height representation
was initially introduced in the context of triangular lattice frustrated Ising
model~\cite{height_ising}, and it also applies for the dimer models via the spin-to-dimer mapping.

In the height representation, each hexagonal plaquette is assigned
an integer number $\tilde{h}(\mathbf{r})$. Turning clockwise around
a site of the even, the height $\tilde{h}(\mathbf{r})$ changes by
$+1$ when crossing an empty link~(or unfrustrated bond), 
and by $-2$ when crossing an occupied link~(or frustrated bond). 

\begin{figure}[t!]
\centering \includegraphics[width=0.3\linewidth]{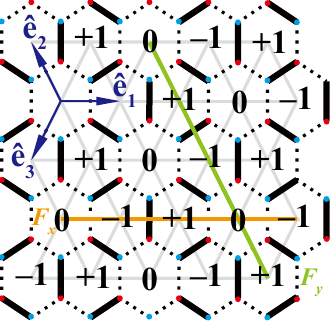} \caption{An illustration of the height field. The red and blue points denote
two sublattices on the dual lattice. The numbers in
the center of hexagonal plaquettes illustrate the height field of
the corresponding dimer configuration. The dark blue arrows denote
the three unit vectors defined on the triangular lattice and its dual
honeycomb lattice of dimers. The light orange and green paths denote
the loop along which the winding numbers $F_{x}$ and $F_{y}$ are
defined. }
\label{fig:s0} 
\end{figure}

\section{Effective field theory near the RK point}
The most interesting physics of the QDM lies around the RK point. Here the RK point serve as a quantum critical point separating the clock and stripe ordered phases. The effective field theory in the vicinity of the RK point takes the form~\cite{Henley,qdm_vbs,height_ising}
\begin{equation}
\mathscr{L}_{0}=\frac{1}{2}\left[(\partial_{\tau}h)^{2}+\kappa^{2}(\nabla^{2}h)^{2}\right]+\frac{\rho_{2}}{2}(\nabla h)^{2}+\lambda\cos2\pi h,\label{eq:L0}
\end{equation}
where $h$ is the coarse-grained height field, $\rho_{2}\propto1-v/t$ controls the phase transition, and $\lambda$ is the instanton term dictating the discrete nature of the height. When $\rho_{2}>0$, the instanton term $\lambda$ is relevant so that the system is pinned to the six-fold clock state; When $\rho_{2}<0$, the fluctuations of $\nabla h$ immediately become unbounded and saturates to its UV limit, where we obtain the staggered stripe state. 
The instanton term $\lambda$ becomes dangerously irrelevant at the critical point $\rho_{2}=0$, which corresponds to the RK point. The effective theory at RK point takes the form of the quantum Lifshitz model
\begin{equation}
\mathscr{L}_{\mathrm{QLM}}=\frac{1}{2}\left[(\partial_{\tau}h)^{2}+\kappa^{2}(\nabla^{2}h)^{2}\right]\label{eq:LQLM}
\end{equation}
with dynamical exponent $z=2$. The fluctuating gauge field mediates short-range interactions between test charges, free from the confinement issue of pure $(2+1)\mathrm{d}$ $U$(1) gauge theories~\cite{Polyakov}.

More recently, the fate of RK point under generic perturbations was discussed in refs.~\cite{rk_2dbp,ashvin1}. On the honeycomb lattice, RK point also admit a relevant trigonal anisotropy $\mathscr{L}_{1}=g_{3}\prod_{\alpha=1}^{3}\left(\nabla h\cdot\hat{\mathbf{e}}_{\alpha}\right)$
in addition to a marginally irrelevant quartic coupling $\mathscr{L}_{2}=g_{4}\left(\nabla h\cdot\nabla h\right)^{2}$. Here $\hat{\mathbf{e}}_{\alpha}$ ($\alpha=1,2,3$) are the unit vectors aligned perpendicular to three dimer directions. Considering the whole theory
\begin{equation}
\mathscr{L}=\mathscr{L}_{0}+\mathscr{L}_{1}+\mathscr{L}_{2},\label{eq:L012}
\end{equation}
it is shown that a sequence of commensurate and incommensurate regime with finite and non-saturated $f$ is stabilized in the vicinity of the RK point. This intermediate regime forms an incomplete `devil's staircase' structure, with the analogue to the fractal `Cantor set' in mathematics. The incommensurate states host gapless phason mode that corresponds to the $U$(1) gauge photons, hence prevents gauge charges from confining. This scenario is dubbed as `Cantor deconfinement' as is proposed in the seminal work~\cite{rk_2dbp}.

\section{Finite size scaling of the multicritical point}

\begin{figure}[ht]
\centering \includegraphics[width=0.4\linewidth]{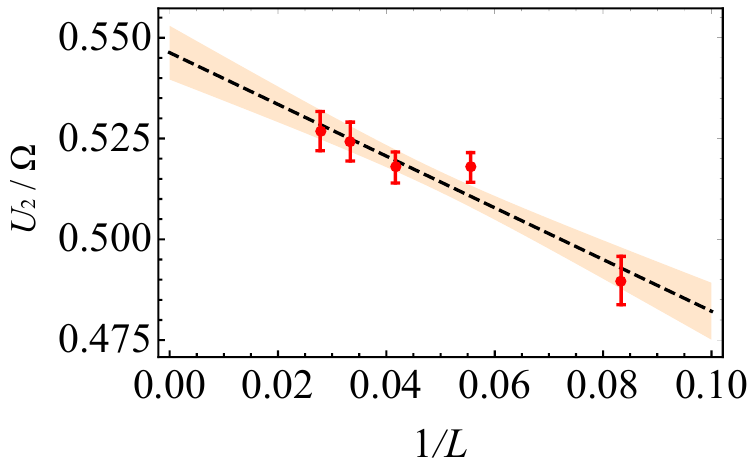} \includegraphics[width=0.4\linewidth]{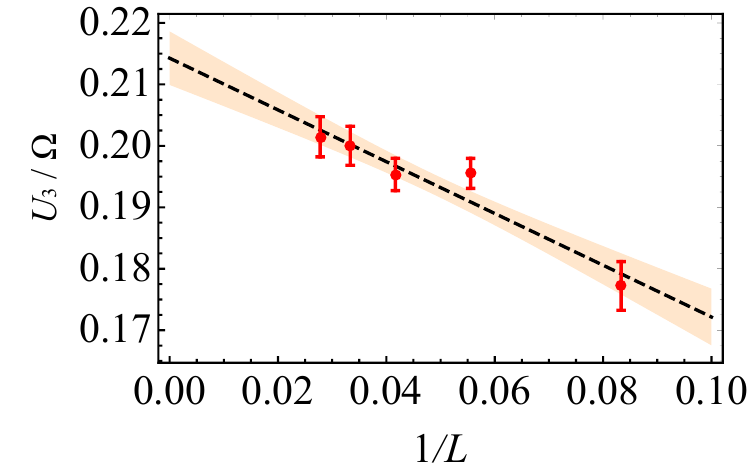}
\caption{The finite-size scaling which gives the position of the multicritical 
point in the thermodynamic limit.}
\label{fig:s5} 
\end{figure}

The phase diagram in the main text is simulated with system size $L=24$.
To determine the location of multicritical point in the thermodynamic limit,
we carried out a finite-size scaling with system sizes $L=12,18,24,30$ and $36$. The position of multicritical point 
is determined by the intersection of clock-stripe transition line and the
stripe-intermediate transition line, \textit{i.e.} $E(f=0)=E(f=2)=E(f=2-3/L_{x})$.
Through an extrapolation, we determine the position of the multicritical point
in the thermodynamic limit, which is $U_{2}/\Omega=0.547(5)$, $U_3/\Omega=0.215(3)$
(Fig. \ref{fig:s5}).

\section{Correlations at the multicritical point}

\begin{figure}[ht]
\includegraphics[width=0.6\linewidth]{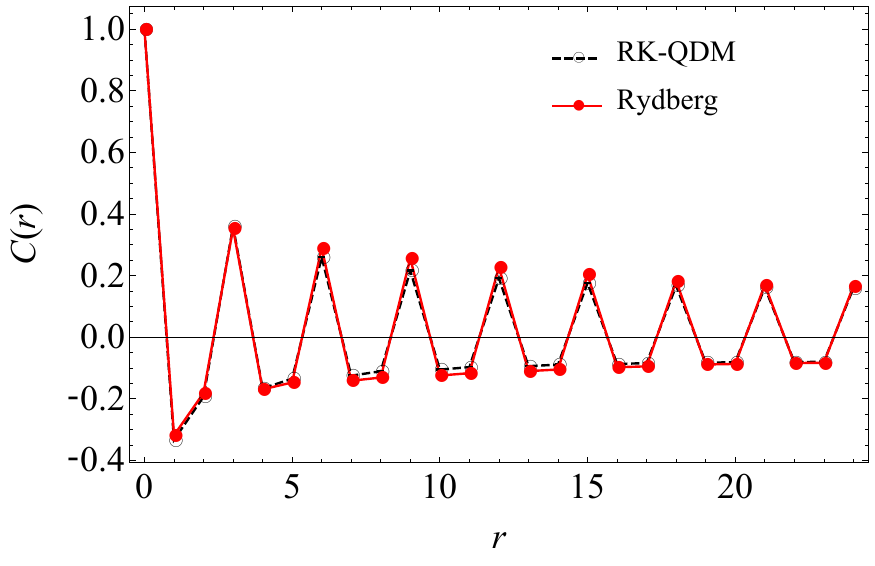} \caption{The spin-spin correlation function along $x$-axis, measured for
the Rydberg model at the multicritical point (solid red)
and the RK-QDM at the RK point (dashed black). }
\label{fig:s2} 
\end{figure}

We calculate the Rydberg density correlation at the multicritical
point (Fig. \ref{fig:s2}). As for comparison, we measure the same correlator 
for the RK wavefunction of the RK-QDM. We restrict our update in Monte Carlo
 simulation to the topological sector $f=0$. We find that these two correlation
functions agrees considerably well, which suggests the RK property
of the multicritical point.

Here we evaluate the long-distance behaviors of order parameter correlators $C_{E}$
and $C_{R}$ at RK point. The asymptotic behaviors of correlators can be directly evaluated from field theory. Here we present
a more intuitive way to understand the long-distance behaviors. The
RK wave function of the RK-QDM is the equal-weight superposition of
fully-packed dimer coverings $|\psi_{\mathrm{RK}}\rangle\sim\sum_{c}|c\rangle$.
The observable $\hat{\mathscr{O}}$ with respect to this wave function
$\langle\psi_{\mathrm{RK}}|\hat{\mathscr{O}}|\psi_{\mathrm{RK}}\rangle$
is equivalent to the statistical problem in classical dimer model
at infinite temperature $T=\infty$, where the statistical weight
of all dimer coverings are identical.

For the electric field correlator $C_{E}$, such statistical
problem has already been solved in the context of dimer models~\cite{fisher1963statistical}.
The asymptotic behavior of the electric field correlator is predicted to be $C_{E}(\mathbf{R})\sim|\mathbf{R}|^{-2}$.
For the Rydberg density correlator $C_R$, such statistical problem have been investigated in the context of classical antiferromagnetic Ising model on the triangular lattice~\cite{dimer_cor}, and the asymptotic
behavior is predicted to be $C_{R}(\mathbf{r}_{i}-\mathbf{r}_{j})\sim|\mathbf{r}_{i}-\mathbf{r}_{j}|^{-1/2}$.

\section{Linear-quadratic crossover}

\begin{figure}[ht]
\centering \includegraphics[width=0.5\linewidth]{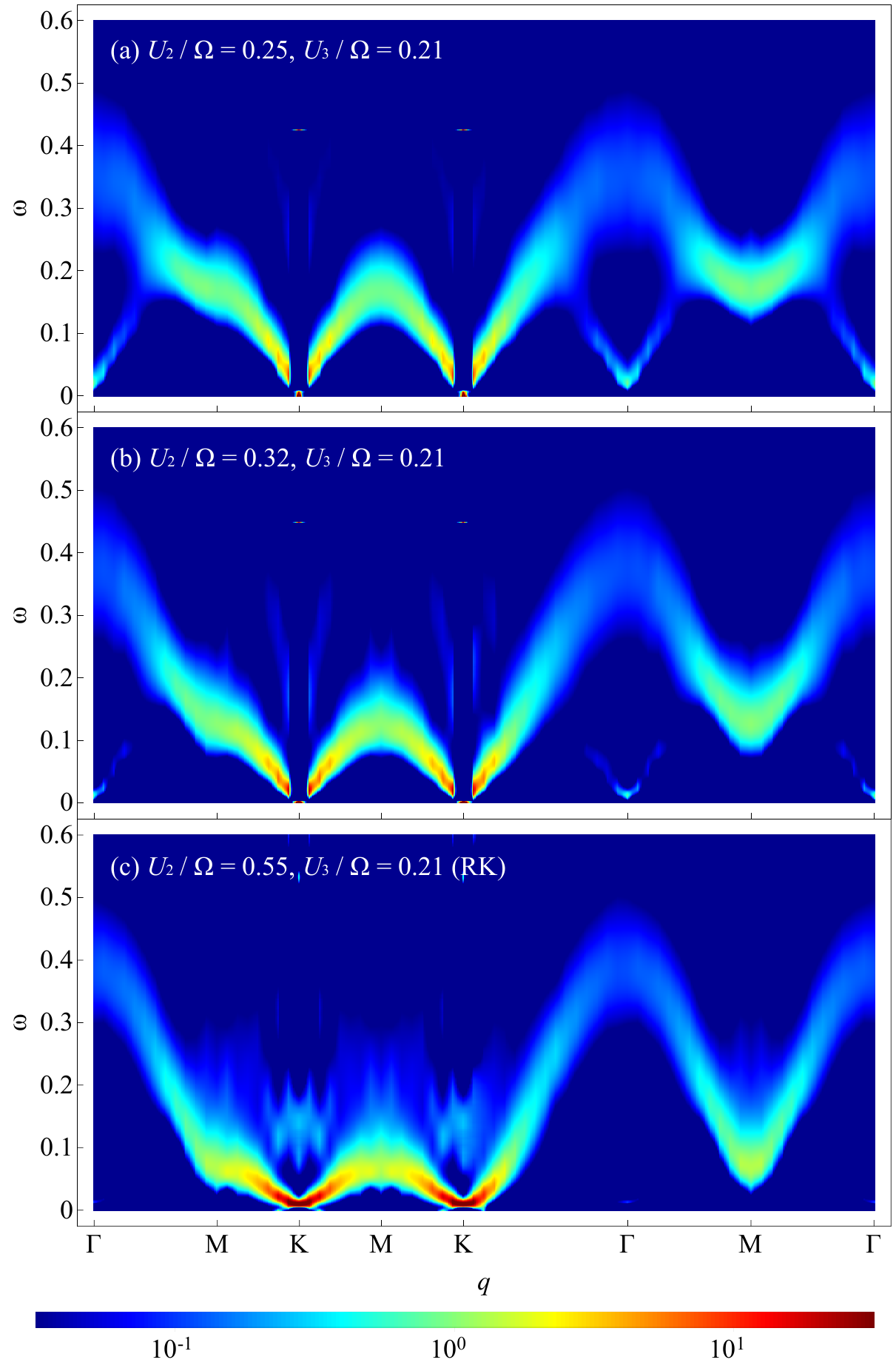} \caption{The linear-quadratic crossover in spin spectrum, measured at $U_{2}/\Omega=0.20$ (\textbf{a}), $U_{2}/\Omega=0.40$ (\textbf{b}), $U_{2}/\Omega=0.55$ (\textbf{c}) (multicritical point) and $U_{3}/\Omega=0.21$.
 In all cases $U_1=5\Omega$, $L=24$, and $\beta=U_1L^{2}/2=288U_1$.}
\label{fig:s4} 
\end{figure}

When far away from the multicritical point, two linear excitations
are found at $\Gamma$ and $K$. When approaching the multicritical
point, the dynamical exponent $z$ changes from $z=1$ to $z=2$.
In the linear-quadratic crossover process (Fig. \ref{fig:s4}), the
linear mode at $K$ softens into a quadratic mode, whereas the mode
at $\Gamma$ gradually vanishes.

\section{Curvatures of quadratic dispersions}

To compare different excitations at the multicritical point, we extract
dispersions of three quadratic excitations and fit them to the form
\begin{equation}
\omega=\frac{1}{2}C_{2}(\mathbf{q}-\mathbf{q}_{0})^{2}\label{eq:c2}
\end{equation}
in vicinity of the gapless momenta $\mathbf{q}_{0}$. 
In addition, we have extracted the corresponding curvatures of the quadratic excitations for the RK-QDM~\cite{rk_debut} at $J=V=1$.
The results are shown in Table~\ref{tbl:1}. We find the curvature of the pi0n{*}
excitation in Rydberg density spectrum roughly twice the curvature of pi0n in
electric field spectrum.

\begin{table}[ht]
\caption{The curvature defined in Eq.~\eqref{eq:c2} for different excitations
measured in the Rydberg model and RK-QDM. }
\begin{tabular}{cc|cc}
\hline 
\multicolumn{2}{c|}{Excitations} & \multicolumn{2}{c}{Curvature $C_{2}$}\tabularnewline
\multicolumn{2}{c|}{} & Rydberg model & RK-QDM\tabularnewline
\hline 
\multirow{2}{*}{(\emph{Electric field})} & resonon & $0.080(3)$ & $0.61(3)$\tabularnewline
 & pi0n & $0.057(4)$ & $0.36(4)$\tabularnewline
(\emph{Rydberg density}) & pi0n{*} & $0.095(2)$ & $0.78(8)$\tabularnewline
\hline 
\end{tabular}\label{tbl:1} 
\end{table}

\section{The intermediate regime}

In the phase diagram, we find an interesting fan-shaped regime with intermediate `tilt' $f$ at the clock-to-stripe boundary. Depending on the value of $f$, the spin structural factor is peaked at some intermediate wave vectors between the K and M points.  The most exciting feature of this intermediate regime is that it meets the microscopic ingredients for the long sought `Cantor deconfinement' scenario~\cite{rk_2dbp}. Due to the limited system size in our measurement, we are unable to observe the fractal structure of this regime.
A detailed analysis of this regime is beyond the scope of this work. However, we notice a numerical work on a similar model with similar phase diagram~\cite{qdm_honey}, where the nature of the intermediate regime and its properties have been elaborately discussed in that work~\cite{qdm_honey}.

\section{Stochastic series expansion (SSE)}

For the numerical works in this paper, we use a QMC method with stochastic series expansion (SSE) algorithm \cite{qmc_1,qmc_2,qmc_3}
to simulate the ground state properties and imaginary time Green
function. This method will be briefly introduced below. For convenience, this part is best described in the effective spin-1/2 representation, by identifying the atomic Rydberg $|\uparrow\rangle$ and ground state $|\downarrow\rangle$ as spin-up and spin-down, so that the Rydberg model maps to the TFIM in the absence of longitudinal magnetic field:
\begin{equation}
    H=\sum_{ij}U_{ij}S_i^zS_j^z-\Omega\sum_iS_i^x.
    \label{eq:tfim}
\end{equation}

In quantum statistics, the measurement of observables is closely related
to the calculation of partition function $Z$ 
\begin{equation}
\langle\mathscr{O}\rangle=\mathrm{tr}\:\left(\mathscr{O}\exp(-\beta H)\right)/Z,\quad Z=\mathrm{tr}\exp(-\beta H)
\end{equation}
where $\beta=1/T$ is the inverse temperature, $H$ is the Hamiltonian
of the system and $\mathscr{O}$ is an arbitrary observable. Typically,
in order to evaluate the ground state property, one takes a sufficiently
large $\beta$ such that $\beta\sim L^{z}$, where $L$ is the system
scale and $z$ is the dynamical exponent. In SSE, such evaluation
of $Z$ is done by a Taylor expansion of the exponential and the trace
is taken by summing over a complete set of suitably-chosen basis.
\begin{equation}
Z=\sum_{\alpha}\sum_{n=0}^{\infty}\frac{\beta^{n}}{n!}\langle\alpha|(-H)^{n}|\alpha\rangle
\end{equation}
We then write the Hamiltonian as the sum of a set of operators whose
matrix elements are easy to calculate. 
\begin{equation}
H=-\sum_{i}H_{i}
\end{equation}
In practice we truncate the Taylor expansion at a sufficiently large
cutoff $M$ and it is convenient to fix the sequence length by introducing
in identity operator $H_{0}=1$ to fill in all the empty positions
despite it is not part of the Hamiltonian. 
\begin{equation}
(-H)^{n}=\sum_{\{i_{p}\}}\prod_{p=1}^{n}H_{i_{p}}=\sum_{\{i_{p}\}}\frac{(M-n)!n!}{M!}\prod_{p=1}^{n}H_{i_{p}}
\end{equation}
and 
\begin{equation}
Z=\sum_{\alpha}\sum_{\{i_{p}\}}\beta^{n}\frac{(M-n)!}{M!}\langle\alpha|\prod_{p=1}^{n}H_{i_{p}}|\alpha\rangle
\end{equation}

To the carry out the summation, a Monte Carlo procedure can be used
to sample the operator sequence $\{i_{p}\}$ and the trial state $\alpha$
with according to their relative weight 
\begin{equation}
W(\alpha,\{i_{p}\})=\beta^{n}\frac{(M-n)!}{M!}\langle\alpha|\prod_{p=1}^{n}H_{i_{p}}|\alpha\rangle
\end{equation}
For sampling we adopt a Metropolis algorithm where the configuration
of one step is generated based on updating the configuration of the
former step and the update is accepted at a probability 
\begin{equation}
P(\alpha,\{i_{p}\}\rightarrow\alpha',\{i'_{p}\})=\min\left(1,\frac{W(\alpha',\{i'_{p}\})}{W(\alpha,\{i_{p}\})}\right)
\end{equation}
Diagonal update, where diagonal operators are inserted into and removed
from the operator sequence, and cluster update, where diagonal and
off-diagonal operates convert into each other, are adopted in update
strategy.

\begin{figure}[b]
\centering \includegraphics[width=0.6\linewidth]{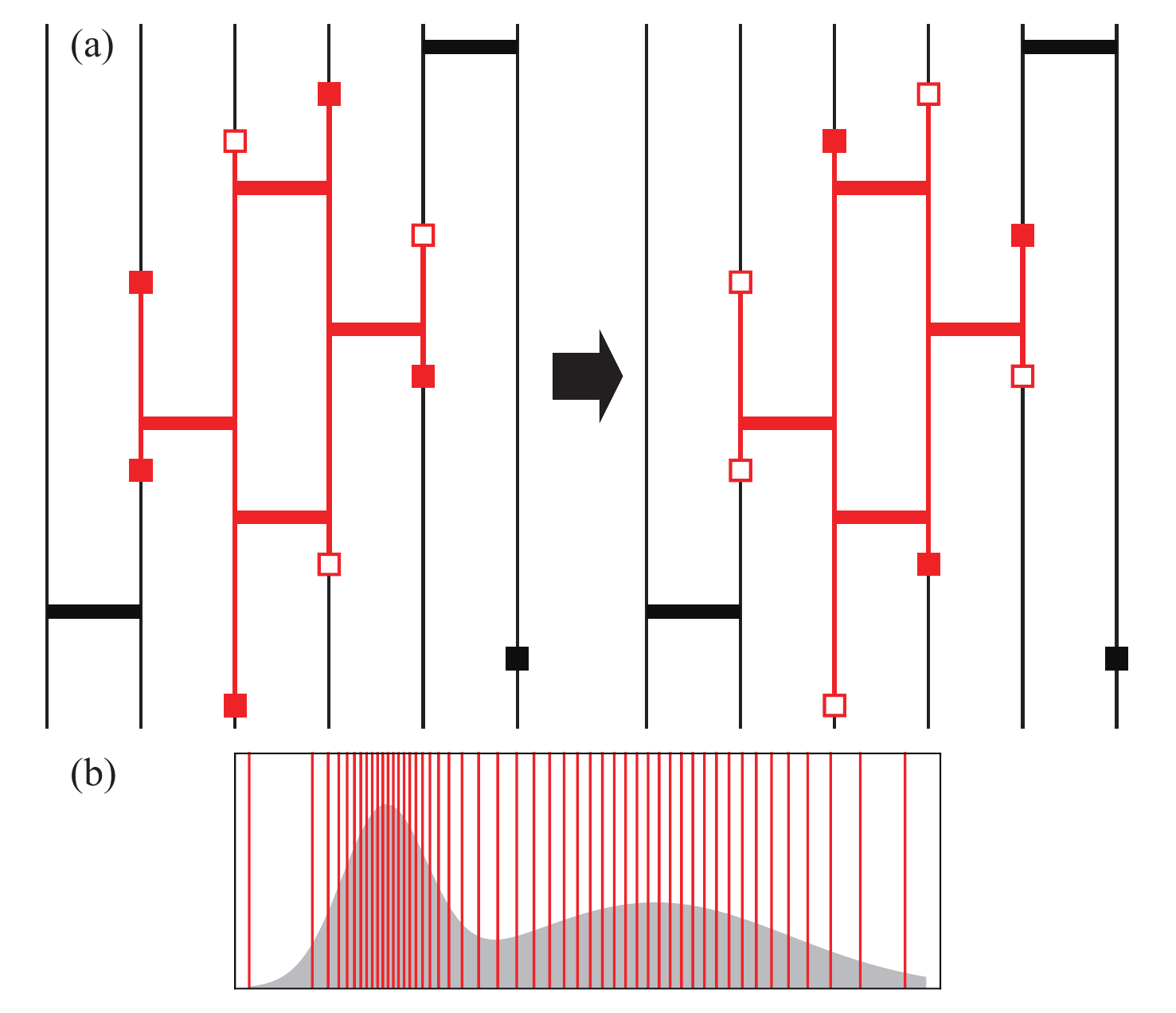} \caption{\textbf{a}, An illustration of the SSE cluster update process where the cluster
marked in red is identified and flipped as a whole. A vertical line
shows a spin expanded. The solid and empty squares show the diagonal
and off-diagonal site operators. The solid bars show the diagonal
bond operators. \textbf{b}, An illustration of the parametrization of a continuous
spectral function into discrete $\delta$-functions. }
\label{fig:s6} 
\end{figure}

In TFIM $H=J\sum_{b}S_{i_{b}}^{z}S_{j_{b}}^{z}-h\sum_{i}\sigma_{i}^{x}$,
we write the Hamiltonian as the sum of following operators 
\begin{equation}
\begin{aligned}H_{0} & =1\\
H_{i} & =h(S_{i}^{+}+S_{i}^{-})/2\\
H_{i+n} & =h/2\\
H_{b+2n} & =J(1/4-S_{i_{b}}^{z}S_{j_{b}}^{z})
\end{aligned}
\end{equation}
where a constant is added into the Hamiltonian for convenience. For
the non-local update, a branching cluster update strategy is constructed
\cite{qmc_2}, where a cluster is formed in $(D+1)$-dimensional by
grouping spins and operators together. Each cluster terminates on
site operators and includes bond operators (Fig. \ref{fig:s6}a).
All the spins in each cluster is flipped together at a probability
$1/2$ after all clusters are identified.

During the simulation process, a number of Monte Carlo steps, denoted $N_t$ are first used for thermalization, i.e., to let the system reach equilibrium. Then, $N_m$ steps are used for measurement. These steps are divided into $N_b$ bins, each containing $N_0=N_m/N_b$ steps. For some given observable $\mathscr{O}$, we first calculate its mean value within each bin $\bar{\mathscr{O}}_b=\frac{1}{N_0}\sum_{i=1}^{N_0}\mathscr{O}_i$, where $b=1,\cdots,N_b$ denotes the numbering of the bin and $i$ denotes the numbering of step within the bin. Then the expected value and the error bar of the observable are given by 
\begin{equation}
    \begin{aligned}
        \langle\mathscr{O}\rangle&=\frac{1}{N_b}\sum_{b=1}^{N_b}\bar{\mathscr{O}}_b\\
        \sigma_\mathscr{O}&=\sqrt{\frac{1}{N_b-1}\left(\langle\mathscr{O}^2\rangle-\langle\mathscr{O}\rangle^2\right)}
    \end{aligned}
\end{equation}

\section{Stochastic analytical continuation (SAC)}

For the spectra in this paper we adopted a stochastic analytical continuation
(SAC) \cite{sac_1,sac_2,sac_3} method to obtain the spectral function
$S(\omega)$ from the imaginary time correlation $G(\tau)$ measured
from QMC, which is generally believed a numerically unstable problem.
This method will be briefly introduced below.

The spectral function $S(\omega)$ is connected to the imaginary time
Green's function $G(\tau)$ through an integral equation 
\begin{equation}
G(\tau)=\int_{-\infty}^{\infty}\mathrm{d}\omega S(\omega)K(\tau,\omega)
\end{equation}
where $K(\tau,\omega)$ is the kernel function depending on the temperature
and the statistics of the particles. We restrict ourselves to the
case of spin systems and with only positive frequencies in the spectral,
where $K(\tau,\omega)=(e^{-\tau\omega}+e^{-(\beta-\tau)\omega})/\pi$.
To ensure the normalization of spectral function, we further modify
the transformation and come to the following equation : 
\begin{equation}
G(\tau)=\int_{0}^{\infty}\frac{\mathrm{d}\omega}{\pi}\frac{e^{-\tau\omega}+e^{-(\beta-\tau)\omega}}{1+e^{-\beta\omega}}B(\omega)\label{eq:kernal}
\end{equation}
where $B(\omega)=S(\omega)(1+e^{-\beta\omega})$ is the renormalized
spectral function.

In practice, $G(\tau)$ for a set of imaginary time $\tau_{i}(i=1,\cdots N_{\tau})$
is measured in QMC simulation together with the statistical errors.
The renormalized spectral function is parameterized into large number
of equal-amplitude $\delta$-functions whose positions are sampled
(Fig. \ref{fig:s6}b) 
\begin{equation}
B(\omega)=\sum_{i=0}^{N_{\omega}}a_{i}\delta(\omega-\omega_{i})
\end{equation}
Then the fitted Green's functions $\tilde{G}_{i}$ from Eq. \ref{eq:kernal}
and the measured Greens functions $\bar{G}_{i}$ are compared by the
fitting goodness 
\begin{equation}
\chi^{2}=\sum_{i,j=1}^{N_{\tau}}(\tilde{G}_{i}-\bar{G}_{i})(C^{-1})_{ij}(\tilde{G}_{j}-\bar{G}_{j})
\end{equation}
where the covariance matrix is defined as 
\begin{equation}
C_{ij}=\frac{1}{N_{B}(N_{B}-1)}\sum_{b=1}^{N_{B}}(G_{i}^{b}-\bar{G}_{i})(G_{j}^{b}-\bar{G}_{j}),
\end{equation}
with $N_{B}$ the number of bins, the measured Green's functions of
each $G_{i}^{b}$.

A Metropolis process is utilized to update the series in sampling.
The weight for a given spectrum is taken to follow a Boltzmann distribution
\begin{equation}
W(\{a_{i},\omega_{i}\})\sim\exp\left(-\frac{\chi^{2}}{2\Theta}\right)
\end{equation}
with $\Theta$ a virtue temperature to balance the goodness of fitting
$\chi^{2}$ and the smoothness of the spectral function. All the spectral
functions of sampled series $\{a_{i},\omega_{i}\}$ is then averaged
to obtain the spectrum as the final result.

\vspace{1mm}

\end{document}